\documentclass{iopart}
\usepackage{iopams}

\def\RR{{\mathbb R}}

\def\stackunder#1#2{\mathrel{\mathop{#2}\limits_{#1}}}
\catcode`\@=11
\def\Let@{\relax\iffalse{\fi\let\\=\cr\iffalse}\fi}
\def\vspace@{\def\vspace##1{\crcr\noalign{\vskip##1\relax}}}
\def\multilimits@{\bgroup\vspace@\Let@
 \baselineskip\fontdimen10 \scriptfont\tw@
 \advance\baselineskip\fontdimen12 \scriptfont\tw@
 \lineskip\thr@@\fontdimen8 \scriptfont\thr@@
 \lineskiplimit\lineskip
 \vbox\bgroup\ialign\bgroup\hfil$\m@th\scriptstyle{##}$\hfil\crcr}
\def\Sb{_\multilimits@}
\def\endSb{\crcr\egroup\egroup\egroup}
\def\Sp{^\multilimits@}

\newcommand{\be}[1]{\begin{equation}\label{#1}}
\newcommand{\ee}{\end{equation}}
\newcommand{\ba}[1]{\begin{eqnarray}\label{#1}}
\newcommand{\ea}{\end{eqnarray}}

\begin{document}

\jl{6}

\title[Multidimensional perfect fluid cosmology]{Multidimensional
perfect fluid cosmology with stable compactified
internal dimensions}

\author{U G\"unther\dag\footnote[3]{e-mail: guenther@pool.hrz.htw-zittau.de},
A Zhuk\dag\ddag\footnote[4]{e-mail: zhuk@paco.odessa.ua}}

\address{\ddag\
Projektgruppe Kosmologie, Institut f\"ur Mathematik,
Universit\"at Potsdam,\\
Am Neuen Palais 10, PF 601533, D-14451 Potsdam, Germany.\\
Fachbereich Physik,
Freie Universit\"at Berlin,\\
Arnimallee 14, D-14195 Berlin, Germany.}

\address{\dag\ Department of Physics, University of Odessa,\\
2 Petra Velikogo St., Odessa 270100, Ukraine.}

\begin{abstract}
Multidimensional cosmological models
in the presence of a
 bare cosmological constant and a perfect fluid
are investigated under dimensional reduction
to $D_0 = 4$ - dimensional effective models.
Stable compactification of the internal spaces is achieved for a
special class of perfect fluids. The external space behaves
in accordance with the standard Friedmann model.
Necessary restrictions on the parameters of the models are found
to ensure  dynamical behaviour of the external (our) universe in
agreement with observations.
\end{abstract}

\pacs{04.50.+h, 98.80.Hw} 

\section{Introduction}

\setcounter{equation}{0}

The large scale dynamics of the observable part of our present time universe
is well described by the Friedmann model with 4-dimensional
Friedmann-Robertson-Walker (FRW) metric. However, it is possible that
space-time at short (Planck) distances might have a dimensionality of more
than four and possess a rather complex topology \cite{1}. String theory \cite
{2} and its recent generalizations --- p-brane, M- and F-theory \cite{3,4}
widely use this concept and give it a new foundation. From this viewpoint,
it is natural to generalize the Friedmann model to multidimensional
cosmological models (MCM) with topology \cite{IMZ} 
\begin{equation}
\label{1.1}M=\RR \times M_0\times M_1\times \ldots \times M_n\ , 
\end{equation}
where for simplicity the $M_i$ \quad $(i=0,\dots ,n)$ can be assumed to be $%
d_i-$dimensional Einstein spaces. $M_0$ usually denotes the $d_0=3$ -
dimensional external space. One of the main problems in MCM consists in the
dynamical process leading from a stage with all dimensions developing on the
same scale to the actual stage of the universe, where we have only four
external dimensions and all internal spaces have to be compactified and
contracted to sufficiently small scales, so that they are aparently
unobservable. To make the internal dimensions unobservable at actual stage
of the universe we have to demand their contraction to scales near to the
Planck length $L_{Pl}\sim 10^{-33}cm$. Obviously, such a compactification
should be stable. Recently \cite{GZ} , we found a class of MCM possessing
stable compactification of extra dimensions.

From the other hand, any realistic MCM should provide a dynamical behaviour
of the external space-time in accordance with the observable universe. The
phe\-no\-me\-no\-lo\-gi\-cal 
approach with a pefrect fluid as a matter source is widely
used in usual 4-dimensional cosmology. According to the present day
observations dynamical behaviour of the universe after inflation is well
described by the standard Friedmann model \cite{Fried} in the presence of a
perfect fluid . Thus it might be worth-while to generalize this approach to
the description of the postinflationary stage in multidimensional
cosmological models. It is desirable to get models where, from one side, the
internal spaces are stably compactified near Planck scales and, from the other
side, the external universe behaves in accordance with the standard
Friedmann model.

Here we present a toy MCM which shows a principal possibility to reach this
goal. This model is out of the scope of MCM with stable compactification
found in \cite{GZ}. The main difference consists in an additional
time-dependent term in the effective potential that provides the needed
dynamical behaviour of the external space-time. This term is induced by a
special type of fine-tuning of the parameters of a multicomponent perfect
fluid. Although such a fine-tuning is a strong restriction on the matter
content of the model, many important cases of physically interest are
described by this class of perfect fluid. We note that a similar class of
perfect fluids was considered in \cite{KZ}, where MCMs were integrated in
the case of an absent cosmological constant and Ricci-flat internal spaces.
As result particular solutions with static internal spaces had been
obtained. According to Sec.4 of the present paper these solutions are not
stable and a bare cosmological constant and internal spaces with
non-vanishing curvature are necessary conditions for their stabilization. In
the present paper we show that with the help of suitably chosen parameters
the model can be further improved to solve two problems simultaneously.
First, the internal spaces undergo stable compactification. Second, the
external space behaves in accordance with the standard Friedmann model.

The paper is organized as follows. In Section~2, the general description of
the considered model is given. In Section~3, the effective potential is
obtained under dimensional reduction to a $D_0$-dimensional (usually $D_0=4$%
) effective theory in the Einstein frame. The problem of stable
compactification is investigated in Section~4 for a toy model with suitably
chosen parameters. Here, it is shown that the external universe behaves as
the standard Friedmann model. Conclusions and references complete the paper.


\section{General description of the model}

\setcounter{equation}{0}

We consider a multidimensional cosmological model on a manifold (\ref{1.1})
in the presence of a perfect fluid and a bare cosmological constant $\Lambda 
$. The metric of the model is parametrized as 
\begin{equation}
\label{2.1}
\fl g=g_{MN}dX^M\otimes dX^N=-\exp{[2\gamma (\tau )]}d\tau \otimes
d\tau +\sum_{i=0}^n\exp{[2\beta ^i(\tau )]}g_{(i)}. 
\end{equation}
Manifolds $M_i$ with the metrics $g_{(i)}$ are Einstein spaces of dimension 
$d_i$, i.e. 
\begin{equation}
\label{2.3}R_{mn}\left[ g^{(i)}\right] =\lambda ^ig_{mn}^{(i)},\qquad
m,n=1,\ldots ,d_i 
\end{equation}
and 
\begin{equation}
\label{2.4}R\left[ g^{(i)}\right] =\lambda ^id_i\equiv R_i. 
\end{equation}
In the case of constant curvature spaces parameters $\lambda ^i$ are
normalized as $\lambda ^i=k_i(d_i-1)$ with $k_i=\pm 1,0$. The scalar
curvature corresponding to the metric (\ref{2.1}) reads 
\begin{equation}
\label{2.4a}\fl
R=\sum_{i=0}^nR_i\exp {\left( -2\beta ^i\right) }+\exp {%
(-2\gamma )}\sum_{i=0}^nd_i\left[ 2{\ddot \beta }^i-2\dot \gamma {\dot \beta 
}^i+{\left( {\dot \beta }^i\right) }^2+{\dot \beta }^i\sum_{j=0}^nd_j{\dot
\beta }^j\right] . 
\end{equation}

Matter fields we take into account in a phenomenological way as a $m-$%
component perfect fluid with energy-momentum tensor 
\begin{equation}
\label{2.4b}T_N^M=\sum_{a=1}^m{T^{(a)}}^M_N , 
\end{equation}
\begin{equation}
\label{2.5}\fl
{T^{(a)}}^M_N = {\rm diag\ } \left( -\rho ^{(a)}(\tau ), 
\underbrace{P_0^{(a)}(\tau ),\ldots ,P_0^{(a)}(\tau )}_{\mbox{$d_0$ times}%
},\ldots , \underbrace{P_n^{(a)}(\tau ),\ldots ,P_n^{(a)}(\tau )}_{%
\mbox{$d_n$ times}}\right) 
\end{equation}
and equations of state 
\begin{equation}
\label{2.8}P_i^{(a)}=\left( \alpha _i^{(a)}-1\right) \rho ^{(a)},\ \ \ \ \
i=0,\ldots ,n,\quad a=1,\ldots ,m. 
\end{equation}
It is easy to see that physical values of $\alpha _i^{(a)}$ according to $%
-\rho ^{(a)}\ \leq \ P_i^{(a)}\ \leq \ \rho ^{(a)}$ run the region $0\leq
\alpha _i^{(a)}\leq \ 2$. The conservation equations we impose on each
component separately 
\begin{equation}
\label{2.6}{T^{(a)}}^M_{N;M} = 0. 
\end{equation}
Denoting by an overdot differentiation with respect to time $\tau ,$ these
equations read for the tensors (\ref{2.5}) 
\begin{equation}
\label{2.7}\dot \rho ^{(a)}+\sum_{i=0}^nd_i\dot \beta ^i\left( \rho
^{(a)}+P_i^{(a)}\right) =0 
\end{equation}
and have according to (\ref{2.8}) the simple integrals 
\begin{equation}
\label{2.9}\rho ^{(a)}(\tau )=A^{(a)}\prod_{i=0}^na_i^{-d_i\alpha _i^{(a)}}, 
\end{equation}
where $a_i \equiv e^{\beta^i}$ are scale factors of $M_i$ and $A^{(a)}$ are
constants of integration. It is not difficult to verify that the Einstein
equations with the energy-momentum tensor (\ref{2.4b})-(\ref{2.9}) are
equivalent to the Euler-Lagrange equations for the Lagrangian \cite{IM},\cite
{Z(QCG)} 
\begin{equation}
\label{2.11}\fl
L=\frac 12e^{-\gamma +\gamma _0}G_{ij}\dot \beta ^i\dot \beta
^j-e^{\gamma +\gamma _0}\left( -\frac 12\sum_{i=0}^nR_ie^{-2\beta ^i}+\kappa
^2\sum_{a=1}^m\rho ^{(a)}+\Lambda \right) . 
\end{equation}
Here we use the notation $\gamma _0=\sum_0^nd_i\beta ^i$ , $\Lambda $ is a
cosmological constant and $\kappa ^2$ is a $D=\sum_0^nd_i\ +1$ - dimensional
gravitational constant. The components of the minisuperspace metric read 
\cite{IMZ} 
\begin{equation}
\label{2.12}G_{ij}=d_i\delta _{ij}-d_id_j \ .
\end{equation}
The Lagrangian (\ref{2.11}) can be obtained by dimensional reduction of the
action 
\begin{equation}
\label{2.13}\fl
S=\frac 1{2\kappa ^2}\int\limits_Md^Dx\sqrt{|g|}\left\{
R[g]-2\Lambda \right\} -\int\limits_Md^Dx\sqrt{|g|}\rho +S_{YGH}=\frac \mu
{\kappa ^2}\int d\tau L. 
\end{equation}
$S_{YGH}$ is the standard York-Gibbons-Hawking boundary term and $\mu
=\prod_{i=0}^nV_i$ , where $V_i$ is the volume of $M_i$ ( with unit scale
factors ) : $V_i={\rm vol\ }(M_i)=\int\limits_{M_i}d^{d_i}y\sqrt{|g^{(i)}|}$
.


\section{The effective potential}

\setcounter{equation}{0}

Let us slightly generalize this model to the inhomogeneous case supposing
that the scale factors $\beta ^i=\beta ^i(x)\ (i=0,\ldots ,n)$ are functions
of the coordinates $x$, where $x$ are defined on the $D_0=(1+d_0)$ -
dimensional external space-time manifold $\bar M_0=\RR \times M_0$ with the
metric 
\begin{equation}
\label{3.1}\bar g^{(0)}=\bar g_{\mu \nu }^{(0)}dx^\mu \otimes dx^\nu
=-e^{2\gamma }d\tau ^2+e^{2\beta ^0(x)}g^{(0)}\ . 
\end{equation}
After conformal transformation of the external space-time metric from the
Brans-Dicke to the Einstein frame: 
\begin{eqnarray}
\label{3.2}
g=g_{MN}dX^M\otimes dX^N & = & \bar g^{(0)}+\sum_{i=1}^n\exp{[2\beta
^i(x)]}g^{(i)} \nonumber \\ 
& = & \Omega ^2\tilde g^{(0)}+\sum_{i=1}^n\exp{[2\beta ^i(x)]}%
g^{(i)}, 
\end{eqnarray}
where 
\begin{equation}
\label{3.3}\Omega ^2={\left( \prod_{i=1}^ne^{d_i\beta ^i}\right) }^{-\frac
2{D_0-2}}, 
\end{equation}
the dimensionally reduced action (\ref{2.13}) reads 
\begin{equation}
\label{3.4}\fl
S=\frac 1{2\kappa _0^2}\int\limits_{\bar M_0}d^{D_0}x\sqrt{%
|\tilde g^{(0)}|}\left\{ \tilde R\left[ \tilde g^{(0)}\right] -\bar
G_{ij}\tilde g^{(0)\mu \nu }\partial _\mu \beta ^i\,\partial _\nu \beta
^j-2U_{eff}\right\} , 
\end{equation}
where $\kappa _0^2=\kappa ^2/V_I$ is the $D_0$-dimensional gravitational
constant, $V_I=\prod_{i=1}^nV_i$ , $\bar G_{ij}\ $ is the midisuperspace
metric with the components 
\begin{equation}
\label{3.5}\bar G_{ij}=d_i\delta _{ij}+\frac 1{D_0-2}d_id_j\ ,\quad
i,j=1,\ldots ,n\quad 
\end{equation}
and the effective potential $U_{eff}$ reads 
\begin{equation}
\label{3.6}U_{eff}={\left( \prod_{i=1}^ne^{d_i\beta ^i}\right) }^{-\frac
2{D_0-2}}\left[ -\frac 12\sum_{i=1}^nR_ie^{-2\beta ^i}+\Lambda +\kappa
^2\sum_{a=1}^m\rho ^{(a)}\right] . 
\end{equation}
The effective action (\ref{3.4}) has the form of a usual 4-dimensional (if $%
d_0=3$) theory and describes a self-gravitating $\sigma -$model with
self-interaction. The internal scale factors play the role of scalar fields
(dilatons in the starting Brans-Dicke frame) satisfying the wave equation 
\begin{equation}
\label{3.6a}\bar G_{ij}\opensquare \beta ^j\equiv \frac 1{\sqrt{|\tilde g^{(0)}|}%
}\partial _\mu \left( \sqrt{|\tilde g^{(0)}|}\bar G_{ij}\tilde g^{(0)\mu \nu
}\partial _\nu \beta ^j\right) =\frac{\partial U_{eff}}{\partial \beta ^i}\
. 
\end{equation}
In the Einstein frame the theory assumes the most natural form \cite{Cho}, 
\cite{LSGAD} and beginning from this point the external space-time metric $%
\tilde g^{(0)}$ is considered as the physical one. For this metric we adopt
following ansatz: 
\begin{equation}
\label{3.7}\tilde g^{(0)}=\Omega ^{-2}\bar g^{(0)}=\tilde g_{\mu \nu
}^{(0)}dx^\mu \otimes dx^\nu =-e^{2\tilde \gamma }d\tilde \tau \otimes
d\tilde \tau +e^{2\tilde \beta ^0(x)}g^{(0)}\ . 
\end{equation}
Thus external scale factors in the Brans-Dicke frame $a_0=e^{\beta ^0}\equiv
a$ and in the Einstein frame $\tilde a_0=e^{\tilde \beta ^0}\equiv \tilde a$
are connected with each other by the relation 
\begin{equation}
\label{3.8}a={\left( \prod_{i=1}^ne^{d_i\beta ^i}\right) }^{-\frac
1{D_0-2}}\tilde a. 
\end{equation}
The energy densities $\rho ^{(a)}$ of the perfect fluid components are given
by  \eref{2.9} and with the help of relation (\ref{3.8}) can be
rewritten as 
\begin{equation}
\label{3.9}\rho ^{(a)}=\rho _0^{(a)}\prod_{i=1}^na_i^{-\xi _i^{(a)}}, 
\end{equation}
where 
\begin{equation}
\label{3.10}\rho _0^{(a)}=A^{(a)}\frac 1{\tilde a^{\alpha _0^{(a)}d_0}} 
\end{equation}
and 
\begin{equation}
\label{3.11}\xi _i^{(a)}=d_i\left( \alpha _i^{(a)}-\frac{\alpha _0^{(a)}d_0}{%
d_0-1}\right) . 
\end{equation}

In the case of one internal space ($n=1$) the action and the effective
potential are respectively 
\begin{equation}
\label{3.12}S=\frac 1{2\kappa _0^2}\int\limits_{\bar M_0}d^{D_0}x\sqrt{%
|\tilde g^{(0)}|}\left\{ \tilde R\left[ \tilde g^{(0)}\right] -\tilde
g^{(0)\mu \nu }\partial _\mu \varphi \,\partial _\nu \varphi
-2U_{eff}\right\} 
\end{equation}
and 
\begin{equation}
\label{3.13}\fl
U_{eff}=e^{2\varphi {\left[ \frac{d_1}{(D-2)(D_0-2)}\right] }%
^{1/2}}\left[ -\frac 12R_1e^{2\varphi {\left[ \frac{D_0-2}{d_1(D-2)}\right] }%
^{1/2}}+\Lambda +\kappa ^2\rho (\tilde a,\varphi )\right] \ ,
\end{equation}
where we redefined the dilaton field as 
\begin{equation}
\label{3.14}\varphi \equiv -\sqrt{\frac{d_1(D-2)}{D_0-2}}\beta ^1\ .
\end{equation}

Let us split the scalar fields $\beta ^i(x)$ in equationss 
(\ref{3.4}) and (\ref{3.6}) in a background component 
$\bar \beta ^i(x)$ and a small
perturbational (fluctuation) component $\eta ^i(x)$ 
\begin{equation}
\label{3.15}\beta ^i(x)=\bar \beta ^i(x)+\eta ^i(x).
\end{equation}
Assuming that such a splitting procedure is well defined we get the
corresponding equations of motion from (\ref{3.6a}) as 
\begin{equation}
\label{3.16}\opensquare \bar \beta ^i=\left[ \bar G^{-1}\right] ^{ij}b_j(\bar \beta
)\ 
\end{equation}
and 
\begin{equation}
\label{3.17}\opensquare \eta ^i=\left[ \bar G^{-1}\right] ^{ij}a_{jk}(\bar \beta
)\eta ^k\ ,
\end{equation}
where 
\begin{equation}
\label{3.18}a_{ij}:=\frac{\partial ^2U_{eff}}{\partial \beta ^i\partial
\beta ^j},\quad b_i:=\frac{\partial U_{eff}}{\partial \beta ^i}\ .
\end{equation}
With the help of an appropriate background depending $SO(n)-$rotation $%
S=S(\bar \beta )$ we can diagonalize matrix $\left[ \bar G^{-1}A\right]
_k^i\equiv \left[ \bar G^{-1}\right] ^{ij}a_{jk}(\bar \beta )$ and rewrite
\eref{3.17} in terms of normal modes $\psi =S^{-1}\eta :$%
\begin{equation}
\label{3.18a}\tilde g^{(0)\mu \nu }D_\mu D_\nu \psi =S^{-1}\bar G^{-1}AS\psi 
\stackrel{def}{=}M^2\psi ,
\end{equation}
where $M^2$ is a background depending diagonal mass matrix 
\begin{equation}
\label{3.18a1}
M^2=diag\left[
m_1^2(\bar \beta ),\ldots  ,m_n^2(\bar \beta )\right] .
\end{equation}
  $D_\mu $ denotes
a covariant derivative 
\begin{equation}
\label{3.18b}D_\mu :=\partial _\mu +\Gamma _\mu +A_\mu \ ,\quad A_\mu
:=S^{-1}\partial _\mu S
\end{equation}
with $\Gamma _\mu +A_\mu $ as connection
on the fibre bundle $E(\bar M_0,\RR^{D_0}\oplus \RR^n)$
consisting of the base
manifold $\bar M_0$ and vector spaces $\RR_x^{D_0}\oplus \RR_x^n=T_x\bar
M_0\oplus \left\{ \left( \eta ^1(x),\ldots ,\eta ^n(x)\right) \right\} $
as fibres.   So, the background components $\bar \beta ^i(x)$
via the effective potential $U_{eff}$ and its Hessian $a_{ij}$ play the role
of a medium for the fluctuational components $\psi ^i(x).$ Propagating in 
$\bar M_0$ filled with this medium  the excitational modes (gravitational
excitons \cite{GZ}) change their masses as well as the direction of 
their "polarization"
defined by the unit vector in the fibre space 
\begin{equation}
\label{3.19}\xi (x):=\frac{\psi (x)}{\left| \psi (x)\right| }\in
S^{n-1}\subset \RR^n,
\end{equation}
where $S^{n-1}$ denotes the $(n-1)-$dimensional sphere.
For considerations on interactions of gravitational excitons
with gauge fields and
corresponding observable effects we refer to \cite{GZ3}. 

We note
that in the general case, when $m^2_i(\bar \beta )\neq m^2_j(\bar \beta ) \
, \ \ i\neq j,$
due to the lack of $SO(n)-$invariance of \eref{3.18a} the
connection $A_\mu $ itself
cannot be interpreted as a $SO(n)-$gauge connection in pure gauge.
This is only possible for $M^2=m^2_{exci}I_n$, with $I_n$ the unit matrix.
Then a transformation
\begin{equation}
\label{3.19a}
\begin{array}{lrcl}
U: \ \ \ & \psi & \mapsto & \tilde \psi =U\psi \\
 & A_{\mu } & \mapsto & \tilde A_{\mu }=UA_{\mu }U^{-1}-(\partial _\mu U)U^{-1}\\
 & D_{\mu } & \mapsto & \tilde D_{\mu }=\partial _\mu +\Gamma _\mu +\tilde A_\mu \\
 & D_{\mu }\psi & \mapsto & \tilde D_{\mu }\tilde \psi =UD_{\mu } \psi \\
\end{array}
\end{equation}
leaves \eref{3.18a} invariant
due to $M^2 \mapsto \tilde M^2=UM^2U^{-1}=M^2$, 
and $U$ is indeed a gauge transformation.

Further from \eref{3.18a} it is clear that a consideration of the
excitational modes  makes only sense if the characteristic space-time scales 
$L_{\bar \beta }$ and $L_\psi $ of the variations of the background fields $%
\bar \beta ^i$ and the excitons $\psi ^i$ are of different order: $L_{\bar
\beta }\gg L_\psi $ . (Otherwise non-perturbative techniques should be
applied.) Coverring the external space-time with domains $\Omega _c$ of
intermedium characteristic length $L_c\approx \left| \Omega _c\right|
^{1/(d_0+1)},$ $L_{\bar \beta }\gg L_c\gg L_\psi $ we can in a crude
approximation replace the background fields $\bar \beta ^i(x)$ in $\Omega _c$
by constants $\bar \beta _c^i$ . According to (\ref{3.16}), (\ref{3.18a})
and due to the
regularity of the minisuperspace metric $\bar G_{ij}$ this implies an
extremum condition on the effective potential in $\Omega _c$%
\begin{equation}
\label{3.20}\left. \frac{\partial U_{eff}}{\partial \beta ^i}\right| _{\vec
\beta _c}=0,
\end{equation}
as well as a vanishing connection $A_{\mu }=0$ and 
the constancy of matrix $M^2$.
The only extremum that provides the constancy of $\bar \beta _c^i$ under
perturbations $\psi ^i$ is a minimum and the exciton masses must be
non-negative $m_{(c)i}^2:=m_i^2(\bar \beta _c)\geq 0$ with at least one of
them strictly positive. (The case of $m_{(c)i}^2=0,\ m_{(c)j}^2>0$ for some $%
i,j$ corresponds to degenerate minima, as e.g. given in Sombrero-like
potentials. The massless modes are similar to Goldstone bosons.)

Models with constant background fields on $\Omega _c=\bar M_0$ and with
effective potentials $U_{eff}$ depending only on the internal scale factors
have been considered in \cite{GZ,GZ2}. The corresponding action functional
reads in this case: 
\begin{eqnarray}\label{3.21}
S & = & \frac{1}{2\kappa ^2_0}\int \limits_{\bar M_0}d^{D_0}x\sqrt {|\tilde
g^{(0)}|}\left\{\tilde R\left[\tilde g^{(0)}\right]
- 2\Lambda _{(c)eff}\right\} + \nonumber\\
& + & \sum_{i=1}^{n}\frac{1}{2}\int \limits_{\bar M_0}d^{D_0}x\sqrt {|\tilde
g^{(0)}|}\left\{-\tilde g^{(0)\mu \nu}\psi ^i_{, \mu}\psi ^i_{, \nu} -
m_{(c)i}^2\psi ^i\psi ^i\right\}\ ,
\end{eqnarray}
where the effective cosmological constant $\Lambda _{(c)eff}$ is connected
with the stable compactification position $a_{(c)i}=\exp \bar \beta _c^i$ by
the relation $\Lambda _{(c)eff}\equiv U_{eff}(\vec \beta _c)$ . From a
physical point of view it is clear that the effective potential should
satisfy following conditions: 
\begin{eqnarray}\label{}
(i)\; \qquad a_{(c)i} & \mbox{\small $^{>}_{\sim}$}& L_{Pl}\ ,
\nonumber\\
(ii)\ \ \, \quad m_{(c)i} & \leq & M_{Pl}\ ,\nonumber\\
(iii)\quad \Lambda _{(c)eff} & \rightarrow & 0\ .\nonumber
\end{eqnarray}The first condition expresses the fact that the internal
spaces should be unobservable at the present time and stable against quantum
gravitational fluctuations. This condition ensures the applicability of the
classical gravitational equations near positions of minima of the effective
potential. The second condition means that the curvature of the effective
potential should be less than Planckian one. Of course, gravitational
excitons can be excited at the present time if $m_i\ll M_{Pl}$. The third
condition reflects the fact that the cosmological constant at the present
time is very small: $|\Lambda| \leq 10^{-56}\mbox{cm}^{-2}\approx
10^{-121}\Lambda _{Pl}$ where $\Lambda _{Pl}=L_{Pl}^{-2}$. Strictly
speaking, in the case that the potential has several minima $(c>1)$ we can
demand $a_{(c)i}\sim L_{Pl}$ and $\Lambda _{(c)eff} \rightarrow 0$ only for
one of the minima to which corresponds the present state of the universe.
For all other minima it may be $a_{(c)i}\gg L_{Pl}$ and $|\Lambda
_{(c)eff}|\gg 0$. %
%

\section{The model}

\setcounter{equation}{0}

A general analysis of the internal spaces stable compactification for MCM
with the perfect fluid (\ref{2.9}) is carried out in our paper \cite{GLZ}.
In the present paper we investigate a particular class of effective
potentials (\ref{3.6}) with separating scale factor contributions from
internal and external factor spaces 
\begin{equation}
\label{4.1}U_{eff}=\stackunder{U_{int}}{\underbrace{{\left(
\prod_{i=1}^ne^{d_i\beta ^i}\right) }^{-\frac 2{D_0-2}}\left[ -\frac
12\sum_{i=1}^nR_ie^{-2\beta ^i}+\Lambda \right] }}\ +\stackunder{U_{ext}}{
\underbrace{\kappa ^2\sum_{a=1}^m\rho _0^{(a)}}.} 
\end{equation}
We will show below, that such a separation on the one hand provides a stable
compactification of the internal factor spaces due to a minimum of the first
term $U_{int}=U_{int}(\beta ^1,\ldots ,\beta ^n)$ as well as a dynamical
behaviour of the external factor space
due to $U_{ext}=U_{ext}(\tilde \beta ^0).$ On
the other hand this separation crucially simplifies the calculations and
allows an exact analysis. The price that we have to pay for the separation
is a fine-tuning of the parameters of the multicomponent perfect fluid 
\begin{equation}
\label{4.2} 
\begin{array}{c}
\alpha _0^{(a)}=\frac 2{d_0}+ 
\frac{d_0-1}{d_0}\alpha ^{(a)} \\  \\ 
\alpha _i^{(a)}=\alpha ^{(a)},\quad i=1,\ldots ,n,\quad a=1,\ldots ,m. 
\end{array}
\end{equation}
Only in this case we have 
\begin{equation}
\label{4.3}\xi _i^{(a)}=-\frac{2d_i}{d_0-1} 
\end{equation}
yielding the compensation of the exponential prefactor for the perfect fluid
term in the effective potential (\ref{3.6}). The corresponding components $%
\rho _0^{(a)}$ read, respectively, 
\begin{equation}
\label{4.4}\rho _0^{(a)}=A^{(a)}\frac 1{\tilde a^{2+(d_0-1)\alpha ^{(a)}}}\
. 
\end{equation}
Although the fine-tuning (\ref{4.2}) is a strong restriction, there exist
some important particular models that belong to this class of multicomponent
perfect fluids. For example, if $\alpha ^{(a)}=1$ the a-th component of the
perfect fluid describes radiation in the space $M_0$ and dust in the spaces $%
M_1,\ldots ,M_n$. This kind of perfect fluid satisfies the condition $%
\sum_{i=0}^nd_i\alpha _i^{(a)}=D$ and is called superradiation \cite{LB}. If 
$\alpha ^{(a)}=2$ we obtain the ultra-stiff matter in all $M_i\,(i=0\ldots
,n)$ which is equivalent, e.g., to a massless minimally coupled free scalar
field. In the case $\alpha ^{(a)} =0$ we get the equation of state $%
P^{(a)}_0=\left[ \left(2-d_0\right) /d_0\right] \rho ^{(a)} $ in the
external space $M_0$ which describes a gas of cosmic strings if $d_0=3$ : $%
P^{(a)}=-\frac 13\rho ^{(a)}$ \cite{SP} and vacuum in the internal spaces $%
M_1,\ldots ,M_n$. If $\alpha ^{(a)} =1/2$ and $d_0=3$ we obtain dust in the
external space $M_0$ and a matter with equation of state $P^{(a)}_i=-\frac
12\rho ^{(a)}$ in the internal spaces $M_i,\quad i=1,\ldots ,n$.

Let us first consider the conditions for the existence of a minimum of the
potential $U_{int}(\beta ^1,\ldots ,\beta ^n)$. According to reference 
\cite{GZ2}
potentials $U_{int}$ of type (\ref{4.1}) have a single minimum if the bare
cosmological constant and the curvature scalars of the internal spaces are
negative $R_i,\Lambda <0.$ The scale factors $\left\{ \beta _c^i\right\}
_{i=1}^n$ at the minimum position of the effective potential are connected
by a fine-tuning condition 
\begin{equation}
\label{4.5}\frac{R_i}{d_i}e^{-2\beta _c^i}=\frac{2\Lambda }{D-2}\equiv 
\widetilde{C},\qquad i=1,\ldots ,n 
\end{equation}
and the masses squared of the corresponding gravitational excitons are
degenerate and given as 
\begin{equation}
\label{4.6} 
\begin{array}{ll}
m_1^2=\ldots =m_n^2=m_{exci}^2 & =- 
\frac{4\Lambda }{D-2}\exp \left[ -\frac 2{d_0-1}\sum_{i=1}^nd_i\beta
_c^i\right] \\  &  \\  
& =2\left| \widetilde{C}\right| ^{\frac{D-2}{d_0-1}}\prod_{i=1}^n\left| 
\frac{d_i}{R_i}\right| ^{\frac{d_i}{d_0-1}}. 
\end{array}
\end{equation}
Further it was shown in reference \cite{GZ2} that the value of the potential $%
U_{int}$ at the minimum is connected with the exciton mass by the relation 
\begin{equation}
\label{4.7}\Lambda _{int}:=U_{int}(\beta _c^1,\ldots ,\beta _c^n)=-\frac{%
d_0-1}4m_{exci}^2\ . 
\end{equation}

From equations (\ref{4.5}), (\ref{4.6}) we see that exciton masses and minimum
position $a_{(c)i}=\exp \bar \beta _c^i$ are constants that solely depend on
the value of the bare cosmological constant $\Lambda $, the (constant)
curvature scalars $R_i$ and dimensions $d_i$ of the internal factor spaces.
This means that we have automatically $\Omega _c=\bar M_0$ from the very
onset of the model. Hence the exciton approach in the present linear form
breaks down only when the excitations become too strong so that higher order
terms must be included in the consideration or the phenomenological perfect
fluid approximation itself becomes inapplicable.

Let us now turn to the dynamical behaviour of the external factor space. For
simplicity we consider the zero order approximation, when all excitations
are freezed, in the homogeneous case: $\tilde \gamma =\tilde \gamma (\tilde
\tau )$ and $\tilde \beta =\tilde \beta (\tilde \tau )$. Then the action
functional (\ref{3.21}) with 
\begin{equation}
\label{4.8}\fl
U_{(c)eff} \equiv U_{eff}\left[ \vec \beta _c,\tilde \beta
(\tilde \tau )\right] =U_{int}(\beta _c^1,\ldots ,\beta _c^n)+U_{ext}\left[
\tilde \beta (\tilde \tau )\right] \equiv \Lambda _{int}+\bar \rho _0(\tilde
\tau ) 
\end{equation}
after dimensional reduction reads: 
\begin{eqnarray}
\label{4.9}
\fl S & = & \frac 1{2\kappa _0^2}\int\limits_{\bar M_0}d^{D_0}x
\sqrt{|\tilde g^{(0)}|}\left\{ \tilde R\left[ \tilde g^{(0)}\right]
-2U_{(c)eff}\right\} =\nonumber \\  
\fl & = & \frac{V_0}{2\kappa _0^2}%
\int d\tilde \tau \left\{ e^{\tilde \gamma +d_0\tilde \beta }e^{-2\tilde
\beta }R[g^{(0)}]+d_0(1-d_0)e^{-\tilde \gamma 
+d_0\tilde \beta }\left( \frac{d\tilde \beta }{d\tilde \tau }\right) ^2
\right. \nonumber \\
\fl & & \left. -2e^{\tilde \gamma +d_0\tilde \beta
}\left( \Lambda _{int}+\bar \rho _0\right) \right\} +
\frac{V_0}{2\kappa _0^2}d_0\int d\tilde \tau \frac d{d\tilde \tau }\left(
e^{-\tilde \gamma +d_0\tilde \beta }\frac{d\tilde \beta }{d\tilde \tau }%
\right) ,
\end{eqnarray}
where usually $R[g^{(0)}]=kd_0(d_0-1),\quad k=\pm 1,0$. The constraint
equation $\partial L/\partial \tilde \gamma =0$ in the synchronous time
gauge $\tilde \gamma =0$ yields 
\begin{equation}
\label{4.10}\left( \frac 1{\tilde a}\frac{d\tilde a}{d\tilde t}\right)
^2=-\frac k{\tilde a^2}+\frac 2{d_0(d_0-1)}\left( \Lambda _{int}+\bar \rho
_0(\tilde a)\right) , 
\end{equation}
which results in 
\begin{eqnarray}
\label{4.11}
\fl
\tilde t+const & = & \int
\frac{d\tilde a}{\left[ -k+\frac{2\Lambda _{int}}{d_0(d_0-1)}\tilde a^2+
\frac{2\kappa ^2}{d_0(d_0-1)}\sum_{a=1}^m\frac{A^{(a)}}{\tilde
a^{(d_0-1)\alpha ^{(a)}}}\right] ^{1/2}}\ , \nonumber \\ 
\fl  & = & \int \frac{d\tilde a}{\left[ -k
+\frac{\Lambda _{int}}3\tilde a^2+\frac{\kappa ^2}3\sum_{a=1}^m
\frac{A^{(a)}}{\tilde a^{2\alpha ^{(a)}}}\right] ^{1/2}},
\end{eqnarray}
where in the last line we put $d_0=3$.

Thus in the zero order approximation we arrived at a Friedmann model in the
presence of negative cosmological constant $\Lambda _{int}$ and a
multicomponent perfect fluid. The perfect fluid has the form of a gas of
cosmic strings for $\alpha ^{(a)}=0$, dust for $\alpha ^{(a)}=1/2$ and
radiation for $\alpha ^{(a)}=1$. As $0\le \alpha ^{(a)}\le 2$ , the
cosmological constant plays a role only for large $\tilde a$ and because of
the negative sign of $\Lambda _{int}$ the universe has a turning point at
the maximum of $\tilde a$. To be consistent with present time observation we
should take 
\begin{equation}
\label{4.12}|\Lambda _{int}|\le 10^{-121}\Lambda _{Pl}.
\end{equation}
We note that due to (\ref{4.11}) and in contrast with (\ref{3.21}) the
minimum value $U_{(c)eff}$ of the effective potential in (\ref{4.8}) cannot
be interpreted as a cosmological constant, even as a time dependent one.
Coming back to the gravitational excitons we see that according to (\ref{4.7}%
) the upper bound (\ref{4.12}) on the effective cosmological constant leads
to ultra-light particles with mass $m_{exci}\leq 10^{-60}M_{Pl}\sim
10^{-32}eV$ . This is much less than the cosmic background radiation
temperature at the present time $T_0\sim 10^{-4}eV$. It is clear that such
light particles up to present time behave as radiation and can be taken into
account as an additional term $\rho _r=\frac{\kappa _0^2A_r/3}{\tilde a^2}$
in  \eref{4.11}. It can be easily seen that we reconstruct the standard
scenario if we consider the one-component $(m=1)$ case with $\alpha
^{(1)}=1/2$, $\kappa ^2A^{(1)}\sim 10^{61}$ and $\kappa _0^2A_r\sim 10^{117}$%
. Here we have at early stages a radiation dominated universe and a dust
dominated universe at later stages of its evolution. 

For completeness we
note that via equations (\ref{4.6}) and (\ref{4.7}) the value of 
the effective cosmological
constant has a crucial influence on the relation between the
compactification scales of the internal factor spaces and their dimensions.
In the case of only one internal negative curvature space $%
M_1=H^{d_1}/\Gamma $ with $R_1=-d_1(d_1-1)$ and compactification scale $%
a_{(c)1}=10L_{Pl}$ we have e.g. the relation \cite{GZ2} $\Lambda
_{int}=-(d_1-1)10^{-2(d_1+2)}L_{Pl}$ , so that the bound (\ref{4.12})
implies a dimension of this space of at least $d_1=59$ . Taking instead of
one internal space a set of 2-dimensional  hyperbolic $g-$tori $\left\{
M_i=H^2/\Gamma \right\} _{i=1}^n$ \cite{rey}  with compactification scale $%
a_{(c)i}=10^2L_{Pl}$ it is easy to check \cite{GZ2} that we need at least $%
n=29$ such spaces to satisfy (\ref{4.12}).

Of course, other values of the cosmological constant lead to other exciton
masses and compactification - dimensionality relations. So, it is also
possible to get models with much more heavier gravitational excitons. For $%
\Lambda _{int}=-10^{-8}\Lambda _{Pl}$ we have e.g. $m=10^{-4}M_{Pl}$ and the
excitons are very heavy particles that should be considered as a cold dark
matter. If we take the one-component case $\alpha ^{(1)}=1$ we get at early
times a radiation dominated universe with smooth transition to a cold dark
matter dominated universe at later stages. But for this example it is
necessary to introduce a mechanism that provides a reduction of the huge
cosmological constant to the observable value $10^{-121}\Lambda _{Pl}$.

%

\section{Conclusion}

\setcounter{equation}{0}

In the present paper we considered multidimensional cosmological models
(MCM) with a bare cosmological constant and a perfect fluid as a matter
source. It can be easily seen that there are only two classes of  perfect
fluids with stably compactified internal spaces. These kind of solutions are
of  utmost interest because an absent  time variation of the fundamental
constants in experiments \cite{M,KPW} shows that at the present time the
extra dimensions, if they exist, should be static or nearly static.

The first class (see \cite{GZ}, \cite{GZ2}) consists of  models with $\alpha
_0^{(a)}=0$. It leads to the vacuum equation of state in the external space $%
M_0$. All other $\alpha _i^{(a)}(i=1,\ldots ,n)$ can take arbitrary values.
This model can be used for a phemenological description of a
muitidimensional inflationary universe with smooth transition to a matter
dominated stage.

In the present paper we found a second class of  perfect fluid models with
stable internal spaces. For these models the  stability is induced by a
fine-tuning of the equation of state of the perfect fluid  in the external
and internal spaces (\ref{4.2}). This class includes many important
particular models and allows considerations of  perfect fluids with
different equations of state in the external space, among them also such
that result in a Friedmann-like dynamics. Thus, this class of models can be
applied for the description of the postinflationary stage in
multidimensional cosmology. For the considered models we found necessary
restrictions on the parameters which, from the one hand, ensure stable
compactification of the internal spaces near Planck length and, from the
other hand, guarantee dynamical behaviour of the external (our) universe in
accordance with the standard scenario for the Friedmann model.

This toy model gives a promising example of a multidimensional cosmological
model which is not in contradiction to observations.

\ack

We thank S.Shabanov and A.Schakel for useful discussions during
the preparation of this paper and the 
 Institute of Mathematics of the Potsdam University and the Institute for
Theoretical Physics of the Berlin Free University for hospitality.
U.G. acknowledges financial support from DAAD (Germany) and A.Z. from
DFG, grant 436 UKR 113.


\end{document}